\documentclass[12pt]{iopart}
\usepackage{graphicx}
\usepackage{iopams}

\usepackage{bm}
\usepackage{harvard}
\newcommand{\be}{\begin{equation}} 
\newcommand{\ee}{\end{equation}}   

\usepackage{array}
\newcolumntype{L}[1]{>{\raggedright\let\newline\\\arraybackslash\hspace{0pt}}m{#1}}
\newcolumntype{C}[1]{>{\centering\let\newline\\\arraybackslash\hspace{0pt}}m{#1}}
\newcolumntype{R}[1]{>{\raggedleft\let\newline\\\arraybackslash\hspace{0pt}}m{#1}}

\usepackage{rotating}

\usepackage{color}

\usepackage{soul}

\begin{document}



\title{A conceptual study on real-time adaptive radiation therapy optimization through ultra-fast beamlet control}

\author{Rodney D. Wiersma, Xinmin Liu}

\address{Department of Radiation and Cellular Oncology, The University of Chicago, Chicago, IL 60637-1470, United States of America}
\ead{rwiersma@uchicago.edu}
\vspace{10pt}
\begin{indented}
\item[]August 2019
\end{indented}

\begin{abstract}
A central problem in the field of radiation therapy (RT) is how to optimally deliver dose to a patient in a way that fully accounts for anatomical position changes over time. As current RT is a static process, where beam intensities are calculated before the start of treatment, anatomical deviations can result in poor dose conformity. To overcome these limitations, we present a simulation study on a fully dynamic real-time adaptive radiation therapy (RT-ART) optimization approach that uses ultra-fast beamlet control to dynamically adapt to patient motion in real-time.

A virtual RT-ART machine was simulated with a rapidly rotating linear accelerator (LINAC) source (60 RPM) and a binary 1D multi-leaf collimator (MLC) operating at 100 Hz. If the real-time tracked target motion exceeded a predefined threshold, a time dependent objective function was solved using fast optimization methods to calculate new beamlet intensities that were then delivered to the patient.

To evaluate the approach, system response was analyzed for patient derived continuous drift, step-like, and periodic intra-fractional motion. For each motion type investigated, the RT-ART method was compared against the ideal case with no patient motion (static case) as well as to the case without the use RT-ART. In all cases, isodose lines and dose-volume-histograms (DVH) showed that RT-ART plan quality was approximately the same as the static case, and considerably better than the no RT-ART case.

Based on tests using several different motion types, RT-ART was able to recover dose conformity to the level that it was similar to an ideal RT delivery with no anatomical changes. With continued advances in real-time patient motion tracking and fast computational processes, there is significant potential for the RT-ART optimization process to be realized on next generation RT machines.
\end{abstract}

%
%
\submitto{Biomedical Physics \& Engineering Express}

%
%
%

\section{Introduction}

Radiation therapy is traditionally a static process, whereby dose treatment plans are calculated based on an initial CT scan of the patient anatomy prior to treatment, and then delivered over the course of a number of weeks. The assumption that the patient’s internal anatomy maintains the same position as the initial CT scan over the entire course of treatment was acceptable in the past, where older technologies were less conformal, however, is no longer acceptable with modern RT methods such as intensity modulated radiation therapy (IMRT), volumetric modulated arc therapy (VMAT), and stereotactic body radiation therapy (SBRT) which can tightly conform the radiation dose to the 3D shape of a tumor with approximately 1-2 mm accuracy \cite{intensity2001intensity,otto2008volumetric}. Lung, prostate, pancreas, liver, and other thoracic and abdominal tumors can move as much as 35mm with breathing, rectal filling, intestinal gas, or other types of biological motion \cite{ross1990analysis,davies1994ultrasound,suh2008analysis}.  Numerous studies have shown how such motion can severely compromise the dosimetric quality of RT plans leading to incomplete target irradiation and unwanted exposure of healthy tissue to high levels of radiation resulting in poor tumor control, tissue toxicity, and other serious health complications for the patient \cite{michaelson2008management}.

Significant efforts have been made to reproduce the initial CT scan position as closely as possible through patient preparation protocols and immobilization devices. These include stereotactic frames, abdominal compression devices, styrofoam body cradles, thermoplastic masks, vacuum lock systems, respiratory gating, and other methods. However, the primary clinical limitation of these methods is that they use conventional static RT treatment planning, and thus cannot fully address changes due weight loss, tumor shrinkage, respiratory motion, intestinal gas movement, abdominal bloating, or other anatomical changes. A newer RT technique, known as adaptive radiation therapy (ART), attempts to partly address this issue by performing rapid dose re-planning before the start of each treatment fraction while the patient is on the LINAC table \cite{yan1997adaptive}. Here, a daily volumetric image is taken and deformable image registration (DIR) is performed to correct for anatomical changes that may have occurred since the initial CT scan. A number of studies have evaluated the benefits of ART, indicating improved target coverage and reduced normal tissue toxicity \cite{van2006conventional,kuo2006effect}. However, although current clinical ART methods may be able to account for inter-fractional patient changes, they are unable to account for intra-fractional motion that may take place during radiation delivery.

A wide variety of methods have been proposed for directly addressing such intra-fractional motion. Among the most advanced, is the use of real-time patient motion compensation, where prior works include the use of a robotic arm to move a compact LINAC in sync with a tumor \cite{adler1997cyberknife}, dynamic multi-leaf collimators (MLC) where the beam defining aperture moves with the tumor \cite{keall2001motion,pommer2013dosimetric,falk2010real}, and moving the patient through use of a dynamical treatment stage \cite{d2005real,wiersma2009development,liu2015robotic,belcher2017toward}. However, these methods typically employ a static treatment plan, and do not dynamically adapt this plan to real-time changes in patient anatomy. Such methods have failed to gain widespread clinical traction due to patient safety risks as they blindly move the radiation beam to follow the tumor without consideration of real-time deformable anatomy changes that can move critical healthy tissues, or other organs at risk (OAR), into the path of the radiation beam.

To address this, one method proposed the use of a negative feedback system to perform real-time adaptive motion optimization through dose accumulation tracking and target motion prediction \cite{lu2009real}. Here leaf opening times of upcoming projections for a binary MLC were calculated just before delivery. In order to keep high dynamic response rates, only one projection was optimized at a time resulting in execution times of less than 100ms per projection. However, a primary drawback of the technique was that a limited set of beamlets could only be optimized at any point in time due to slow gantry speeds. Another method was adaptive IMRT sequencing for a MR-linac \cite{kontaxis2015new,kontaxis2015towards,kontaxis2017towards}. The system consists of an iterative sequencing loop open to anatomy updates and an adaptation scheme that enables convergence to the ideal dose distribution. However, it was aimed at conventional linear accelerators with slow gantry speeds, and also restricted dose optimization to a small subset of gantry angles that can compromise overall plan dose optimality in comparison to performing optimization over the entire gantry rotation.

To overcome these limitations, we propose a radical departure from conventional static RT by introducing a fully dynamic optimization strategy that inherently incorporates patient motion. By using patient feedback information,  an optimization problem with a novel time dependent objective function is solved such that a fast rotating radiation source and beamlet control system can dynamically adapt beamlet intensities during delivery. In this scenario each fraction of the RT treatment is delivered as a series of continually adapting micro-fractions where each micro-fraction is dose optimal and temporally short in order to capture patient motion. With continued advances in real-time patient motion tracking \cite{lagendijk2008mri,wiersma2008combined,mutic2014viewray}, fast optimization processes \cite{sempau2000dpm,jia2012gpu}, and fast LINAC gantry/beamlet speeds \cite{fan2013toward} there is significant potential for the RT-ART optimization process to be fully implemented on future RT devices.

\section{Methods}

\subsection{System design}

\begin{figure}
	\centerline{
		\includegraphics[width=150mm]{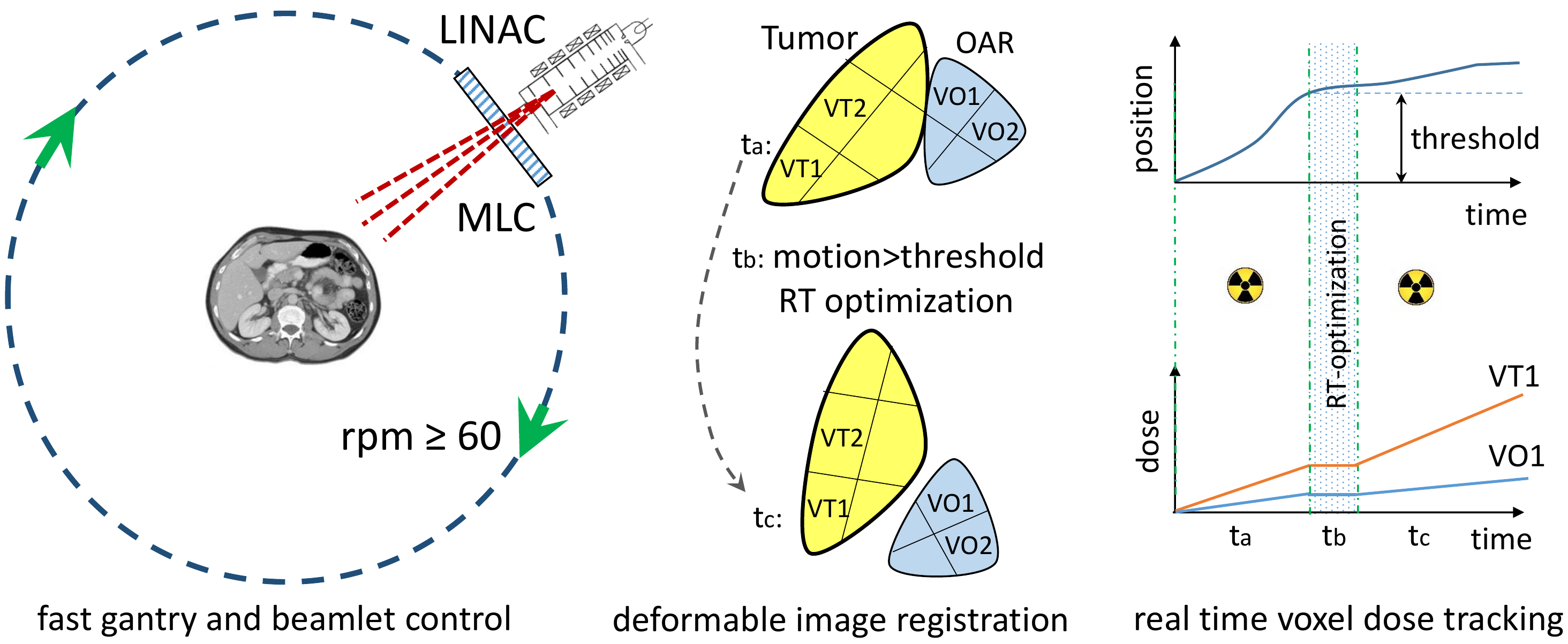}
	}
	\caption{A conceptual dynamical optimization radiation system where a radiation source rapidly rotates around the patient with a binary multi-leaf collimator that allows real-time beamlet control. By real-time tracking of structure volumes the accumulated dose to voxels can be tracked during radiation delivery.}
	\label{fig_system}
\end{figure}

The method is based on all beamlets being accessible in real-time such that on-the-fly dose adaptation can be performed. In the proposed design (Figure~\ref{fig_system}), it was assumed that radiation beamlets were located 360 degrees around the patient in a ring-like configuration such as formed by a radiation source that is rapidly rotated around the patient at 60 RPM \cite{fan2013toward}. The modeled beamlet control method consisted of a binary multi-leaf collimator (MLC) that can be rapidly switched on and off at 100Hz \cite{carol1995peacock,fan2013toward}. It was assumed that a suitable third party imaging system exists that can provide real-time tracking information to the RT-ART algorithm. Development of such a tracking device is beyond the scope of this work, however, potential methods include MV+kV fluoroscopy \cite{wiersma2008combined,grelewicz2014combined}, MRI guidance \cite{lagendijk2008mri,mutic2014viewray}, infrared (IR) markers \cite{wiersma2009development}, 3D surface imaging \cite{wiersma2013spatial}, or other methods. Using such positional information together with the LINAC beam on/off history therefore allows calculation of the cumulative voxel dose history.

\begin{figure}
\centerline{
	\includegraphics[width=80mm]{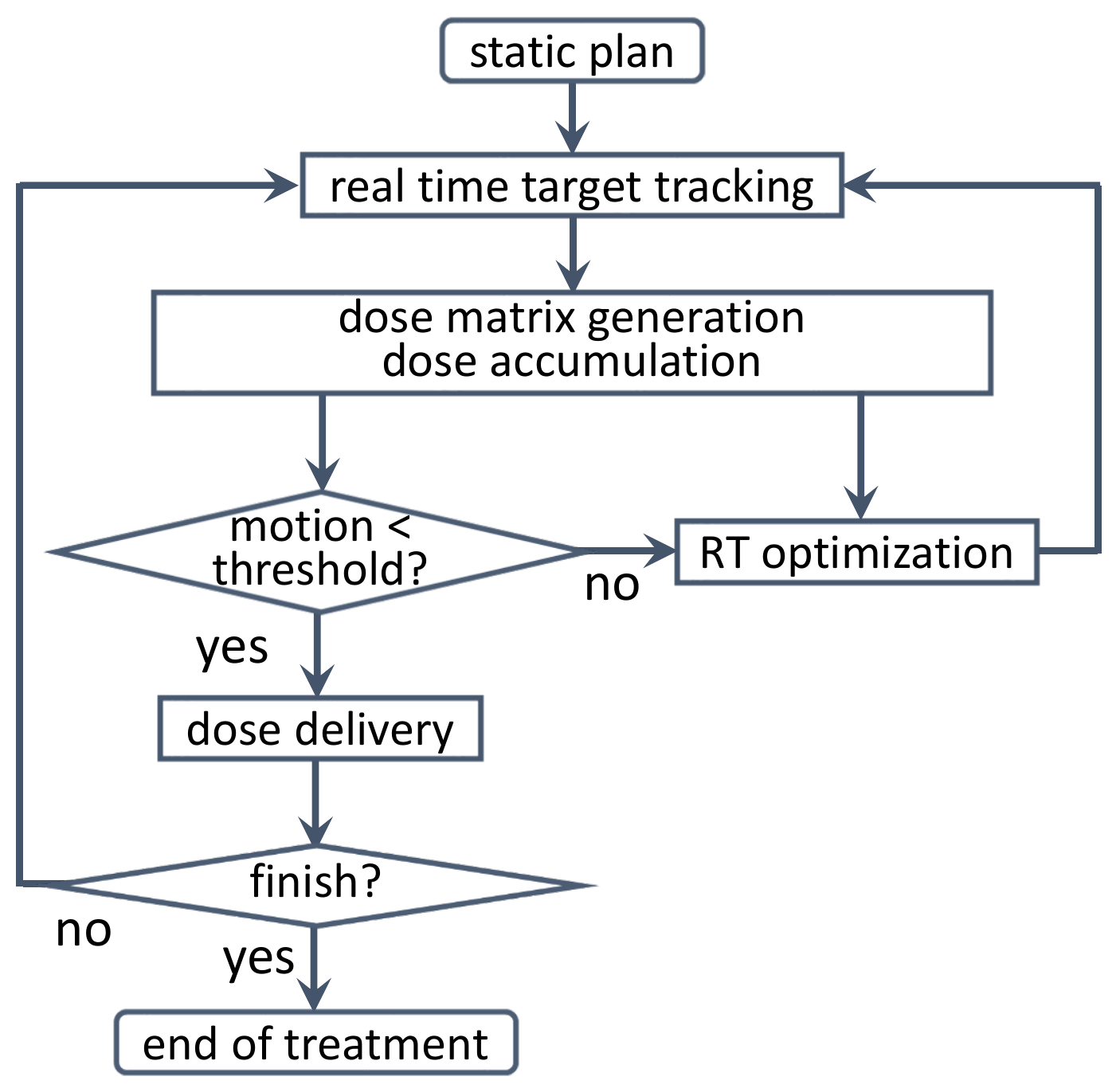} }
	\caption{
		Proposed work flow for a dynamic optimization radiation therapy system.}
	\label{fig_flows}
\end{figure}

The workflow of the proposed RT-ART method is shown Figure~\ref{fig_flows}. Similar to conventional RT, a CT simulation is first acquired and a plan generated before the start of treatment with targets and OARs segmented and satisfying prescribed doses. During treatment, real-time imaging is used to monitor the position of structures and used to judge whether or not motion is within a preset threshold. If the motion is below threshold, RT-ART delivers radiation as according to the initial beamlet intensities. In this case, the treatment is similar to conventional RT. If motion exceeds the threshold, RT-ART generates an updated plan based on the real-time structure positions and current voxel dose history.

\subsection{Real-time beamlet intensity optimization}

With the assumption of sufficiently fast gantry rotation and MLC speeds, treatment planning can be considered a fluence map optimization (FMO) problem, and solved by convex optimization methods. In this work we used the interior point optimizer (IPOPT) \cite{biegler2009large}, although other optimization algorithms such as proximal operator graph solver (POGS) \cite{boyd2011distributed,liu2017use} or quasi-Newton methods such as the limited-memory Broyden-Fletcher-Goldfarb-Shanno (L-BFGS) \cite{morales2011remark} could also be used. The real-time adaptive optimization problem is formulated as,
\begin{equation}
\begin{array}{rl}
\mbox{minimize} & f(d_n-p) \\
\mbox{subject to} & d_n=d_{n-1}+D_n x_n,
                  \quad D_n=h(z(t_n)), \quad x_n\geq 0\\
\end{array}\label{form_ART}
\end{equation}
Here it is important to note that the objective function is dynamic in that it is a function of the real-time patient anatomy position $z(t)$. The function $f$ can be considered as a piecewise quadratic function on dose deviation/overdosing/underdosing, and $p$ are the prescribed dose to the target and the tolerance dose to OARs The dose matrix, $D_n$, relating beamlet to voxel is no longer static, but rather can change as a function $h(z(t_n))$ of updated voxel positions as provided by the real-time tracking system. This is fundamentally different from current IMRT and VMAT methods which are static in nature and thus do not consider time. The treatment starts at the normal position $z(t_0)=0$ with $t_0=0$. The real-time beamlet intensity re-optimization is triggered at moment $t_n$, $n\geq 1$, whenever $|z(t_n)-z(t_{n-1})|\geq $ threshold. The voxel dose accumulation history $d_{n-1}$ is calculated based on $z(t)$, $0\!\leq\! t \! \leq \! t_n$ and LINAC beam on/off information, where

\begin{equation}
d_{-1}=0, \qquad d_{n-1}= d_{n-2} +
\displaystyle\int_{t_{n-1}}^{t_n} h(z(t))\; x_{n-1}\; dt, \quad \mbox{for} \;\; n\geq 1.
\end{equation}

Beamlet intensities $x_{n}$ are optimized as in Eq. \ref{form_ART} by taking into account dose history and minimizing the dose deviation $f$ with respect to the prescribed dose and OAR dose constraints $p$.

A 6 MV LINAC beam was used were the incident fluence was discretized into a rectangular grid of beamlets. The set of beamlets for which dose was calculated was based on an isotropic 2.5 mm expansion of the union of all targets. A beamlet was included in the fluence map if its central axis intersects the enlarged target. The dose matrix $D$ was calculated using a pencil beam algorithm as provided by the matRad open source multi-modality radiation treatment planning system \cite{craft2014shared,cisternas2015matrad}. For dose calculation, the original CT image was down-sampled to a lower resolution, and was gridded in three dimensions. The coordinate system and the conversion of voxel indices to spatial location was given as in \cite{craft2014shared}. Based on the real-time CT scan, the dose matrix $D$ was calculated for the purpose of voxel dose tracking and re-optimization.

\subsection{Motion simulation}

A CT scan following the TG-119 protocol \cite{ezzell2009imrt} was used to simulate input data that would be provided by a hypothetical imaging system capable of real-time tracking of the target structure. A C-shaped target surrounding a central avoidance structure was created. The outer arc of the target is 3.7 cm in radius and approximately 3 cm long. The central OAR is a cylinder 1 cm in radius and approximately 4 cm long. The gap between the target and OAR was 0.5 cm, so the inner arc of the target is 1.5 cm in radius. The total phantom dimension was 19.0 cm $\times$ 13.0 cm $\times$ 9.0 cm.

The dynamical response of the RT-ART system was tested using prior patient recorded lung and prostate tumor motion that was specifically chosen to represent a wide variety of different motion types \cite{wiersma2008use,tehrani2013real}. Specifically, three types of motion were selected: step-like, continuous drift, and periodic. In all cases only rigid-body target motion was simulated by moving the entire target structure within the TG-119 CT.

A prescribed planning target volume (PTV) dose of 50 Gy was used. The static case (no patient motion) was set as the reference standard in which to judge the quality of the RT-ART system. For each motion type investigated, the RT-ART method was compared against the static case as well as the case without the use of RT-ART (patient motion). Isodose lines and dose-volume-histograms (DVH) where used to judge RT-ART plan quality. Other metrics used were PTV-D95, the lowest dose encompassing 95\% of the target volume, and PTV-V95, volume of PTV receiving 95\% of the prescription dose or more.

\section{Results}

\begin{figure*}
	\centerline{
\includegraphics[width=160mm]{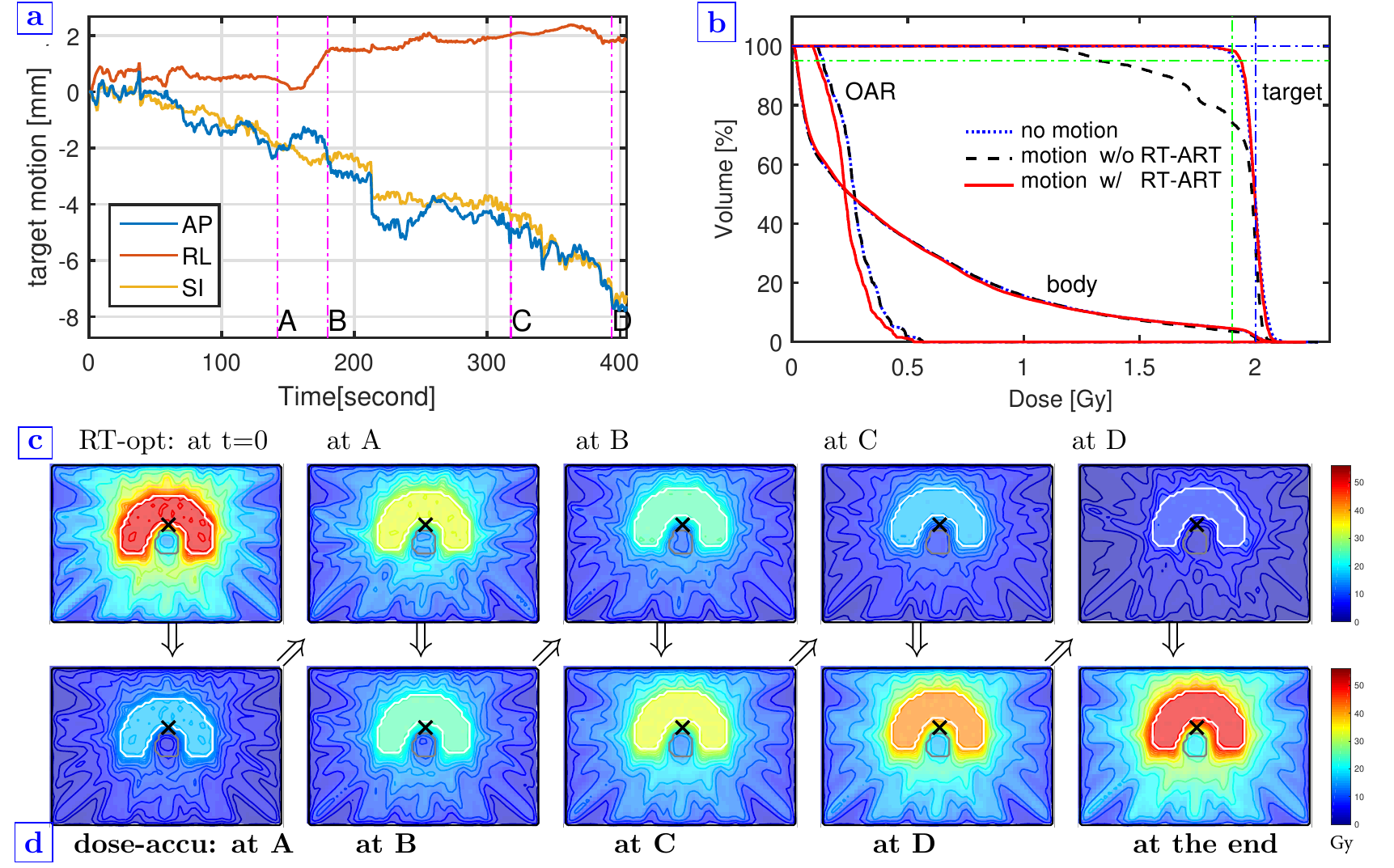}
	}
	\caption{
Application of RT-ART method to intra-fractional continuous drift prostate tumor motion on a TG-119 phantom.
(a) The motion threshold of 3 mm was exceeded at time points t = A, B, C and D resulting in trigging of RT-ART. (b) DVH curves show dose with (solid red line) and without (dashed black line) the use RT-ART, and the case with RT-ART generated DVH curves similar to the static case (dotted blue line).
With RT-ART, it has a significant improvement over the case without RT-ART.
The bottom two rows show the temporal evolution of the planned (c) and delivered (d) over the course of treatment for time points A, B, C, and D.
Arrows indicate how voxel dose accumulation tracking is used as an input in solving the RT-ART optimization problem. The symbol X indicates isocenter.	 }
	\label{fig_tg119_drift_1}
\end{figure*}

The general response of the RT-ART system to patient motion is shown in Figure~\ref{fig_tg119_drift_1}. For continuous drift motion, the predefined 3 mm tolerance level was exceeded 4 times at points A, B, C, and D (Figure~\ref{fig_tg119_drift_1}a). At each point, the RT-ART process was automatically triggered in that the optimization problem as defined in (\ref{form_ART}) was solved using real-time voxel dose accumulation and structure position information. As expected, for a continually adapting system, isodose lines show how over the course of delivery the planned (RT-opt) dose decreases (Figure~\ref{fig_tg119_drift_1}c), whereas, the delivered (accu-dose) dose increases until reaching the prescribed 50 Gy treatment goal (Figure~\ref{fig_tg119_drift_1}d).

DVH curves are shown in Figure~\ref{fig_tg119_drift_1}b. With RT-ART, the generated DVH curves were similar to the static case, whereas, without the use of RT-ART, the dose to the target was severely compromised. The PTV-D95/PTV-V95 was found to be 98.5\%/97.5\% for the RT-ART case and was similar to the case of the static plan with no motion of 98.5\%/98.3\%. Compared to without RT-ART, a significant improvement was seen, where PTV-D95 improved from 74.1\% to 98.5\% and PTV-V95 improved from 67.5\% to 97.5\%.

\begin{figure}
	\includegraphics[width=80mm]{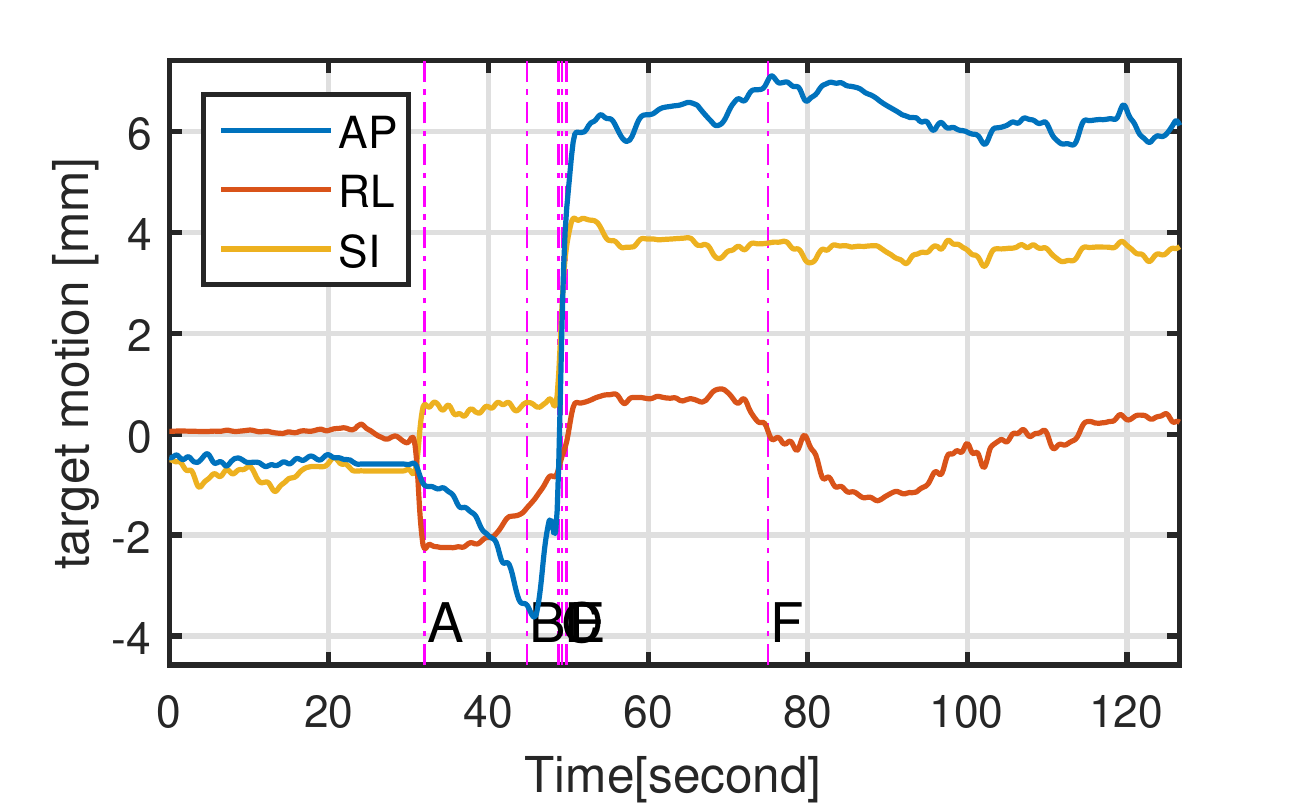}
	\includegraphics[width=80mm]{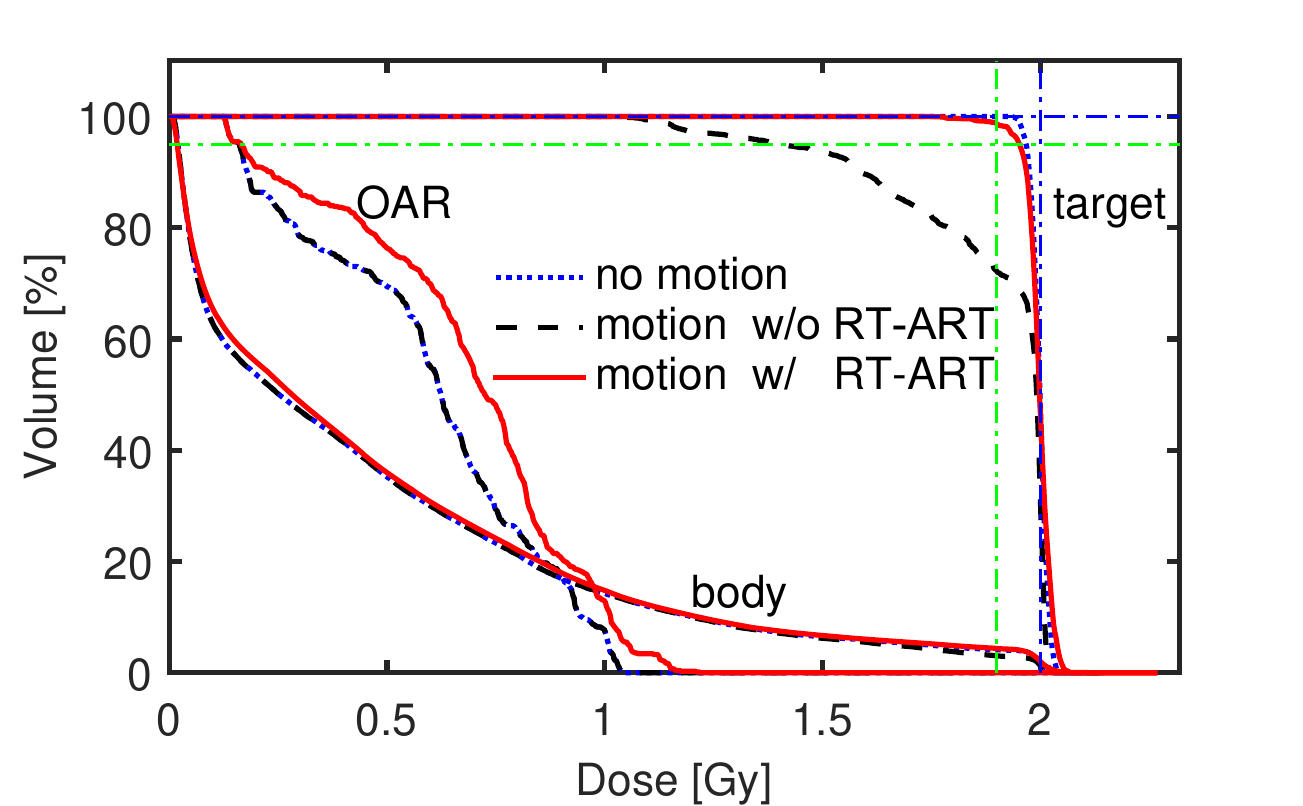}	
	\caption{
Application of RT-ART method to intra-fractional step-like prostate tumor motion on a TG-119 phantom.
DVH curves showing dose delivered to the TG-119 phantom with (solid red line) and without (dashed black line) RT-ART for step-like motion and for a static plan (dotted blue line).
The motion threshold was exceed three times in rapid succession, resulting in re-optimization points C-E falling on top of each other.
	}
	\label{fig_tg119_step_like_2}
\end{figure}

The results of the RT-ART process as applied to step-like motion are summarized in Figure~\ref{fig_tg119_step_like_2}. Similar to the continuous drift motion case, the RT-ART process was triggered whenever target displacement exceeded 3 mm displacement from the previous re-optimization point. Unlike continuous drift motion, where structure volume velocities were slow, the high target velocities around the step lead to three rapid re-optimizations (time points C-E) taking place before target motion stabilization. Comparing the RT-ART case to the static case, dosimetric conformity was found to be approximately the same, whereas, the no RT-ART case showed poor PTV dose coverage. The PTV-D95 was found to be 98.7\% for the RT-ART case and was significantly better than the dynamic case of 72.7\%.
Note that for static plan with no target motion, PTV-D95 was 98.5\% and PTV-D95 was 98.3\% (dotted blue line).

\begin{figure}
	\includegraphics[width=80mm]{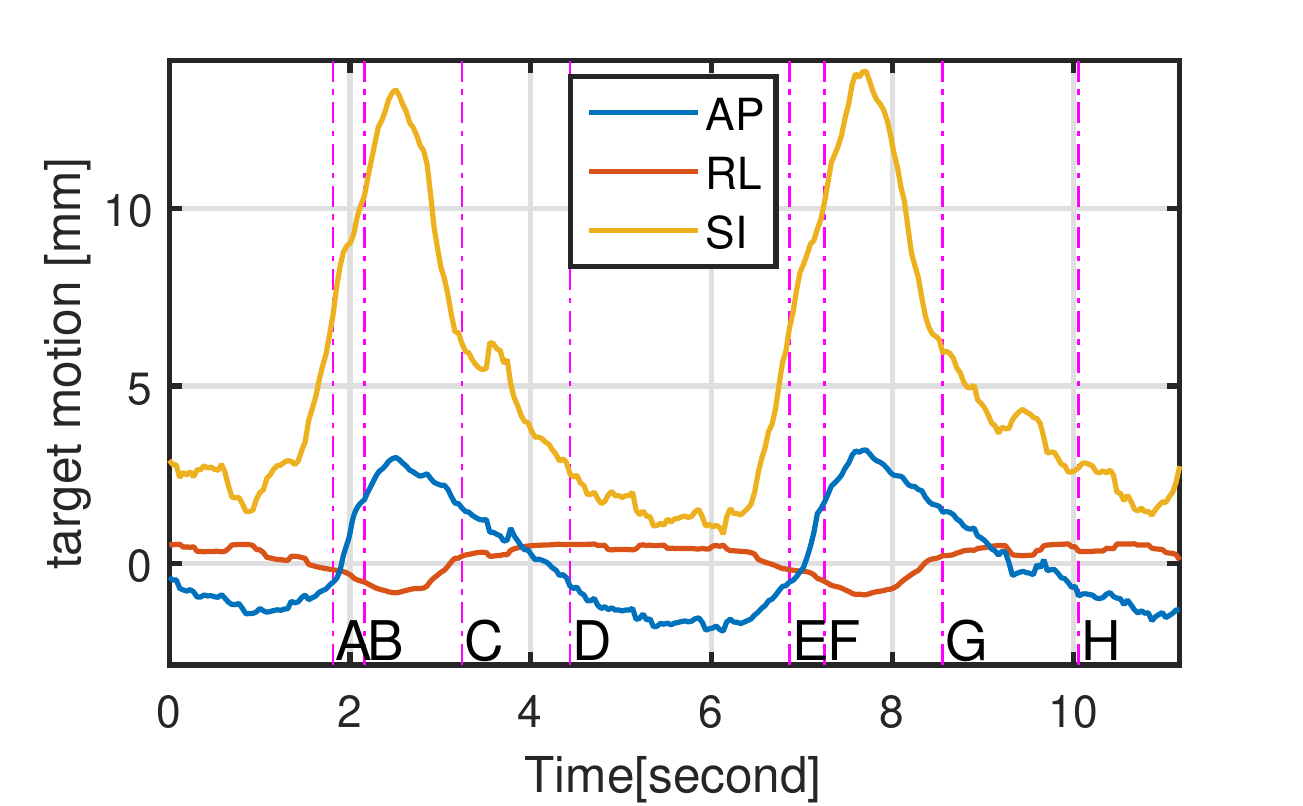}
	\includegraphics[width=80mm]{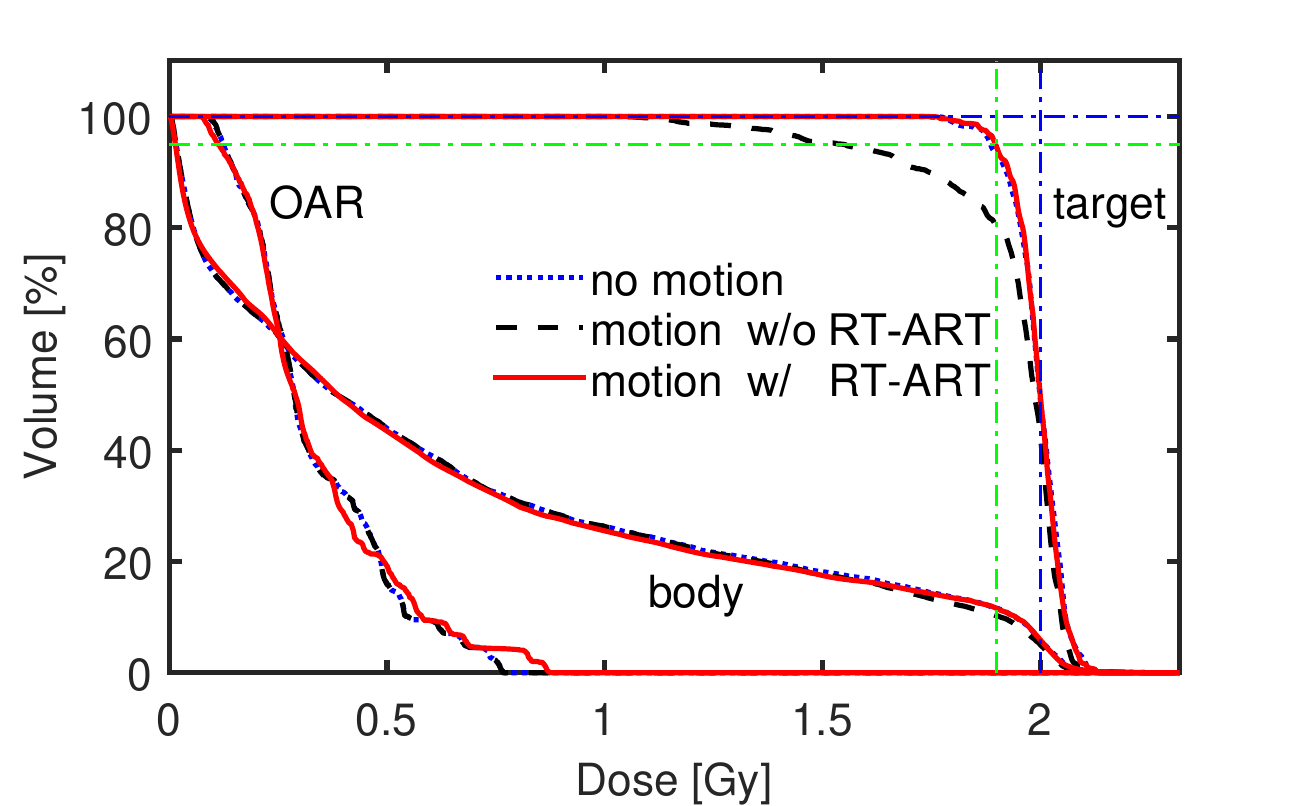}
	\caption{
Application of RT-ART method to intra-fractional periodic lung tumor motion  on a TG-119 phantom. DVH curves showing dose delivered to the TG-119 phantom with (solid red line) and without (dashed black line) the use RT-ART, and  static plan with no target motion (dotted blue line).
	}
	\label{fig_tg119_breath_2}
\end{figure}

The most challenging case was the application of RT-ART to periodic input motion that is typical for tumors undergoing respiratory motion. Due to high target velocities, the tolerance threshold may be exceed multiple times during each respiratory cycle. As shown in Figure~\ref{fig_tg119_breath_2}, the RT-ART re-optimization process was triggered near mid inhale-exhale transition points. Analysis of DVH curves shows similar dosimetric conformity between the RT-ART case and the static case, whereas, the no RT-ART case showed poor dose coverage. The PTV-D95 was found to be 94.5\% for the RT-ART case and was significantly better than without the use of RT-ART (80.4\%).
And the PTV-V95 was 94.9\% and 77.4\% with and without the use RT-ART, respectively.
Note that for static plan with no target motion, PTV-D95 was 93.3\% and PTV-D95 was 94.3\%.

\section{Discussion}

This work aims to establish the fundamental framework for the RT-ART method by solving the time-dependent objective function as defined by Eq. \ref{form_ART}. Using the standard TG-119 treatment planning commissioning phantom and several patient motion types (step-like, continuous drift, and periodic), significant improvements in overall dose conformity were achieved. Across all motion types investigated, the mean PTV-D95/V95 improved from 75.6\%/71.8\% without RT-ART, to 97.2\%/96.7\% with RT-ART. This compares favorably to the mean PTV-D95/V95 of 96.8\%/97.0\% for an ideal no motion treatment.

To demonstrate the core RT-ART optimization framework, a simplistic model where only the target was translated in rigid-body fashion was used. Although adequate for this proof-of-concept study, in actual clinical practice targets and OARs can undergo translations, rotations, and deformations. Here it will be necessary to use real-time deformable image registration (DIR) methods that will allow full tracking of treatment planning volumes. Development of such real-time volumetric tracking systems to provide such input information for RT-ART is beyond the scope of this work, however, it should be noted that recent work in real-time DIR of cine-mode MR have demonstrated sustained performance of 250ms per image, with an accuracy of approximately one voxel \cite{sharp2018real}. Inclusion of such real-time DIR at part of the RT-ART process is currently under investigation.

The hypothetical LINAC assumed in this study had beamlets located around the patient in a ring like configuration and relies on fast gantry rotation and MLC speeds to allow real-time access to beamlets allowing the system to dynamically adapt to patient motion. This configuration is fundamentally different from conventional LINACs, which typically have slow gantry and MLC speeds, that would prevent beamlets from being quickly accessed at real-time speeds. The RT-ART hardware configuration is potentially feasible, as it is similar to how modern CT scanners operate, where slip-ring technology can allow a x-ray source to rotate around the patient at speeds of up to 0.33 second per rotation (180 RPM) \cite{petersilka2008technical}. In one promising LINAC design, gantry rotation and MLC speeds of 1 second per rotation (60 RPM) and 100 Hz beamlet open/close times, respectively, have been demonstrated \cite{fan2013toward}. In another design, 16 stationary LINACS located around the patient are electronically controlled such that the system can switch between beam directions in just 300 ns without requiring gantry rotation \cite{maxim2019phaser}. Such upcoming LINAC designs would ideally fit the RT-ART method, as it would allow fast access to the entire beamlet space.

The RT-ART configuration greatly simplifies optimization, as both gantry and MLC leaf velocity constraints do not exist, the machine produced fluence map would be similar to an ideal fluence map, such that the entire optimization process becomes convex in nature. Such convex problems are easy to solve, and guarantee discovery of a global minimum, resulting in the most optimal dose solution for the patient. This is fundamentally different from conventional LINACs, where gantry and MLC velocity constraints form a non-convex problem that will not guarantee a global minimum \cite{shepard2002direct,otto2008volumetric}.

As the RT-ART process is aimed to dynamically capture patient motion, it is necessary to perform dose calculation and optimization in real-time. Recent research in modern computer algorithms and technologies such as the use GPU parallelization have greatly increased the speed of dose calculation \cite{sempau2000dpm,jia2012gpu}. In one implementation, use of CPU parallelization and vectorization have demonstrated the ability to perform the 4D dose reconstruction in approximately 15 ms \cite{ziegenhein2018real}. Even without the use of hardware parallelization, modern optimization algorithms, such as the proximal operator graph solvers have shown a 1-2 order magnitude speed increase compared to conventional algorithms \cite{liu2017use}. It is envisioned that future advances in algorithms and hardware will further reduce RT-ART processing times making the method clinically feasible.

\section{Conclusion}

The core framework for solving the RT-ART optimization problem was developed and presented for the first time. Based on tests using patient derived step-like, continuous drift, and period motion, the RT-ART method has the potential of addressing patient motion changes without compromising dose conformity. With continued advances in real-time patient motion tracking and fast computational processes, there is significant potential for the RT-ART optimization process to be realized on next generation RT machines.

\section*{References}

\bibliography{references}

\end{document}